\documentclass{article}

\usepackage{microtype}
\usepackage{graphicx}
\usepackage{siunitx}
\usepackage{subfigure}
\usepackage{booktabs} 
\usepackage{amsmath}

\usepackage{hyperref}

\usepackage[accepted]{icml2020}

\icmltitlerunning{Bach or Mock? A Grading Function for Chorales in the Style of J.S. Bach}

\begin{document}

\twocolumn[
\icmltitle{Bach or Mock? A Grading Function for Chorales in the Style of J.S. Bach}




\begin{icmlauthorlist}
\icmlauthor{Alexander Fang}{cs,music}
\icmlauthor{Alisa Liu}{cs}
\icmlauthor{Prem Seetharaman}{cs}
\icmlauthor{Bryan Pardo}{cs}
\end{icmlauthorlist}

\icmlaffiliation{cs}{Department of Computer Science}
\icmlaffiliation{music}{Bienen School of Music, Northwestern University, Evanston, IL, USA}

\icmlcorrespondingauthor{Alexander Fang}{alexanderfang2019@u.northwestern.edu}
\icmlcorrespondingauthor{Alisa Liu}{alisa@u.northwestern.edu}
\icmlcorrespondingauthor{Prem Seetharaman}{prem@u.northwestern.edu}
\icmlcorrespondingauthor{Bryan Pardo}{pardo@northwestern.edu}

\icmlkeywords{automatic evaluation, Bach chorales, music generation, deep learning, domain knowledge}

\vskip 0.3in
]



\printAffiliationsAndNotice{}  

\section{Introduction}\label{sec:introduction}
Deep generative systems that learn probabilistic models from a corpus of existing music do not explicitly encode knowledge of a musical style, compared to traditional rule-based systems \cite{Ebcioglu, nierhaus2009algorithmic}. Thus, it can be difficult to determine whether deep models generate stylistically correct output without expert evaluation. However, human evaluation is expensive and time-consuming, limiting when it can be performed during the research cycle. Moreover, the experimental setup and execution vary greatly across human subject studies, hindering comparability of results. Therefore, there is a need for automatic, interpretable, and musically-motivated evaluation measures of generated music. Such grading functions can allow researchers to efficiently evaluate their models, shed insight into the musical strengths and limitations of generated output, and serve as a consistent benchmark for comparing different models.

In this paper, we introduce a grading function that evaluates four-part chorales in the style of J.S. Bach along important musical features. The Bach chorales represent a canonical dataset for music generation models that has been used in multiple prior works  \cite{Liang2017AutomaticSC, huang2017convolution,pmlr-v70-hadjeres17a}, due to the dataset's size and stylistic consistency. We use the grading function to evaluate the output of a Transformer model, and show that the function is both interpretable and outperforms human experts at discriminating Bach chorales from model-generated ones.


\begin{table*}[t!]
    \fontsize{7.5}{7.2}\selectfont
    \centering
    \caption{The median value (standard deviation) for every feature in the grading function, as well as the overall grade, for Bach chorales and generated chorales. Lower values represent better chorales. We can see that the model struggles with avoiding parallelisms.}
    \vskip 0.1in
    \begin{tabular}{c|ccccccccc|c}
    \toprule
        & Note & Rhythm & \begin{tabular}{@{}c@{}}Parallel \\ Errors\end{tabular} & \begin{tabular}{@{}c@{}}Harmonic \\ Quality\end{tabular} & S Intervals & A Intervals & T Intervals & B Intervals & \begin{tabular}{@{}c@{}}Repeated \\ Sequence \end{tabular}&\begin{tabular}{@{}c@{}}\textbf{Overall} \\ \textbf{Grade} \end{tabular}\\\midrule
        Bach&0.24 (0.15)&0.23 (0.14)&0.0 (0.69)&0.41 (0.2)&0.47 (0.28)&0.49 (0.23)&0.53 (0.24)&0.69 (0.4)&1.29 (0.88)&4.91 (1.63)\\
        Generated&0.37 (0.22)&0.26 (0.14)&2.16 (3.22)&0.54 (0.31)&0.53 (0.35)&0.71 (0.34)&0.73 (0.38)&0.89 (0.68)&1.86 (2.81)&8.94 (4.64)\\\bottomrule
    \end{tabular}
    \label{tab:feature_analysis}
\end{table*}

\section{Grading Function for Four-Part Chorales}\label{sec:grading-function}
Given a four-part chorale, our grading function\footnote{https://github.com/asdfang/constraint-transformer-bach/tree/master/Grader} outputs a real-valued grade. We represent a chorale as a set of distributions, each corresponding to a musical feature important for evaluating Bach-style chorales. We implement our grading function using music21 \cite{cuthbert2010music21}. 


For each feature $f$ (described in Section \ref{subsec:features}), we use the Wasserstein metric \cite{ruschendorf1985wasserstein} to measure the distance between the distribution $P_{c}^f$ of the given chorale $c$ and the ground-truth distribution $P_{\text{Bach}}^f$ over the set of true Bach chorales. By taking a weighted sum of the Wasserstein distances over all the features (Eq. \ref{equation:grading-equation}), we obtain the overall grade for a chorale. Note the output of the grading function is positive, and a lower grade represents a better chorale. 
\begin{equation}\label{equation:grading-equation}
    g(c)=\sum_{f\in\text{features}} w_f\cdot\text{Wass}\left(P_{c}^f, P_{\text{Bach}}^f\right)
\end{equation}


\subsection{Features}\label{subsec:features}
In this section, we describe each feature used to represent a chorale (or set of chorales). The weight $w_f=1$ unless stated otherwise.

\subsubsection{Pitch}
The pitch distribution is the distribution of a chorale's pitches in scale degrees. We consider enharmonic spellings as distinct, but not octave displacements. For a concrete example, if a chorale in C Major had 60 C's, 25 F$\sharp$'s, and 15 G$\flat$'s, the probabilities for $\hat{1}$ (``scale degree $1$''), $\sharp\hat{4}$, and $\flat\hat{5}$ are $.60$, $.25$, and $.15$, respectively. The pitch distribution feature evaluates a Bach-like usage of tonality, distinguishing pieces that are too chromatic (e.g. twelve-tone pieces) and ones that are too stagnant (e.g. never uses \textit{any} chromaticism).



\subsubsection{Rhythm}
The rhythm distribution is the distribution of note lengths in units of quarter notes, e.g. eighth notes are $0.5$ units, quarter notes are $1.0$. This feature serves to measure whether chorales use rhythm like Bach does: eighths and quarters as the main body, and others for decoration and variety. 

\subsubsection{Intervals}
The interval distribution is the distribution of directed melodic interval sizes, i.e. ascending and descending intervals of the same distance are different. Each voice (soprano, alto, tenor, bass) serves a different musical function; specifically, melodies in soprano parts have the most intervallic variety, bass parts leap more frequently for harmony, and tenor and alto parts tend to employ mostly small intervals. Therefore, we measure the interval distribution separately for each voice, for a total of four interval distributions.

\subsubsection{Harmonic qualities}
The harmonic qualities distribution describes the usage of vertical harmony by keeping only the quality, e.g. ``D Major'' would be reduced to ``major.'' This feature also helps encourage a Bach-like usage of 18th century tonality by majority of major, minor, and dominant-seventh chords.

\subsubsection{Parallel errors}
The parallel errors distribution is the distribution of occurrences of the hallmark part-writing errors: parallel fifths and octaves (including unisons) in similar and contrary motion. 
Observe that what matters is not only the distribution between parallel fifths and octaves, but also the count of these errors relative to the total number of notes. Therefore, the Wasserstein distance for this feature is multiplied by $w_{\text{parallel errors}}=\frac{\text{error to note ratio of chorale }c}{\text{error to note ratio of Bach}}$. This weight is large if the given chorale has a large error to note ratio compared to real Bach chorales, thereby penalizing the chorale.

\subsubsection{Repeated sequences}
The repeated sequence distribution is the distribution of the length (in units of quarter notes) of sequences containing at least two notes and appearing at least twice in the chorale, in order to promote a Bach-like handling of recurring motifs and intentional musical repetition. To identify repeated sequences, we use the dynamic programming algorithm in \cite{hsu1998pattern}.



\begin{figure}
    \centering
    \includegraphics[scale=0.175]{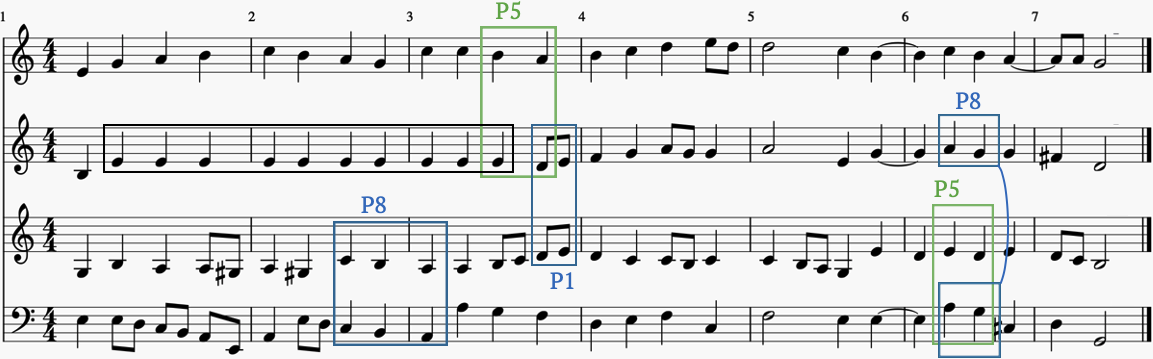}
    \caption{A generated chorale receiving an overall grade of $26.0$ with a parallel error distance of $13.9$ and repeated sequence distance of $5.9$. The features with the largest values represent weaknesses of the composition. In the figure, P1 is parallel unisons, P5 is parallel fifths, and P8 is parallel octaves.}
    \label{fig:mock_score}
\end{figure}

\begin{figure}[t!]
    \centering
    \includegraphics[scale=0.45]{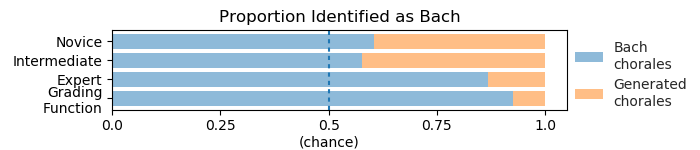}
    \caption{Results of the paired discrimination experiment carried out on human listeners. The grading function ``picks'' the chorale that receives the better grade and achieves $92\%$ accuracy, outperforming human experts at $86.7\%$.}
    \label{fig:human_results}
\end{figure}

\section{Experiments}

We now show that the grading function provides interpretable output and is a promising substitute for human evaluation. We used the grading function to evaluate the output of a Transformer model \cite{NIPS2017_7181} with relative attention \cite{DBLP:journals/corr/abs-1809-04281} trained on a corpus of 351 Bach chorales, using the same data representation as in \cite{pmlr-v70-hadjeres17a}. 

The grade distribution for Bach chorales and generated chorales is very well-separated with a Kolmogorov–Smirnov test p-value of $1$e$-78$. In Table \ref{tab:feature_analysis}, we compare the median value of every feature in the grading function. Generated chorales do worse than Bach chorales in every feature.

To further show the grading function's interpretability, we display a badly graded generated chorale in Figure \ref{fig:mock_score}. We see especially large distances for its parallel error and repeated sequence features. Indeed, the grading function automatically found six total parallel errors and identified an abnormally long sequence of repeated quarter notes (the repeated E's in measures 1--3 of the alto voice).

To compare our grading function to human performance, we performed a paired discrimination test with $n=36$ responses. We assessed the musical expertise of our participants through a series of pre-test questions, and assigned them to one of three groups: novice ($n_0=16$), intermediate ($n_1=15$), and expert ($n_2=5$). In the paired discrimination test, we presented three \textit{pairs} of audio examples representing complete chorales, one Bach and one generated, and asked participants to select the one composed by Bach. In Figure \ref{fig:human_results}, we compare the human pick to selecting the chorale that receives a better grade. We find that the grading function achieves 92.6\% accuracy, outperforming human experts at 86.7\%.

\nocite{allan2004harmonising, Boulanger2012polyphonic}
\bibliography{main}
\bibliographystyle{icml2020}

\end{document}